\documentclass[12pt]{article}
\usepackage{amssymb} 
\usepackage{epsfig}
\usepackage{float}
\newcommand{\be}{\begin{equation}}
\newcommand{\ee}{\end{equation}}
\newcommand{\bea}{\begin{eqnarray}}
\newcommand{\eea}{\end{eqnarray}}

\begin{document} 

\begin{center}
{\bf 
NEUTRINO OSCILLATIONS IN THE FRAMEWORK OF THE  TREE-NEUTRINO MIXING }
\end{center}

\begin{center}
S. M. Bilenky 
\footnote{\em Report at the Ist Yamada Symposium on
                          Neutrinos and Dark Matter in Nuclear Physics,
                                           June 9--14, 2003, Nara, Japan.}
\end{center}
\vspace{0.1cm} 
\begin{center}
{\em  Joint Institute
for Nuclear Research, Dubna, R-141980, Russia\\}
\end{center}
\begin{center}
{\em INFN, Sez. di Torino and Dip. di Fisica Teorica,
Univ. di Torino, I-10125 Torino, Italy\\}
\end{center}

\begin{abstract}
In the framework of the three-neutrino mixing the neutrino oscillations
in the atmospheric and solar ranges of neutrino mass-squared differences 
are considered in the leading approximation. Neutrinoless double $\beta$- decay is also discussed. 
\end{abstract}

\section{Introduction}

Convincing evidence in favor of neutrino oscillations, driven by small neutrino masses and neutrino mixing, were obtained in the 
Super-Kamiokande \cite{S-K},
 SNO \cite{SNO},  KamLAND \cite{Kamland} and other neutrino
experiments \cite{ndm03}. These findings opened a new field of research: physics of massive and mixed neutrinos.

In spite of big progress achieved in the recent years, there are many opened fundamental problems in the field. Some of them we will discuss later.

We will consider  neutrino oscillations and neutrinoless double $\beta$-decay in the framework of the three neutrino mixing \footnote{We will not consider here indications in favor of neutrino oscillations that were obtained in the LSND  experiment. The results of the LSND experiment will be checked by the MiniBOONE experiment  which is  going on now at Fermilab \cite{ndm03}}
Main features of neutrino oscillations in the three-neutrino case are determined by the smallness of two
parameters $\frac{\Delta m^{2}_{21}}{\Delta m^{2}_{32}}$ and $|U_{e3}|^{2}$.
We will consider here transition probabilities in the leading approximation.
As we will see, in this approximation neutrino oscillations in the solar and
atmospheric ranges of $\Delta m^{2}$ are described by two-neutrino type
formulas and are decoupled  (see \cite{BGG})).
 This picture of neutrino oscillations corresponds to existing experimental data.

\section{Neutrino oscillations}

The standard neutrino charged and neutral currents have the form

\begin{equation}
j^{\mathrm{CC}}_{\alpha} =2\, \sum_{l} \bar \nu_{lL} \gamma_{\alpha}l_{L};\,~~
j^{\mathrm{NC}}_{\alpha} =\sum_{l} \bar \nu_{lL}\gamma_{\alpha}\nu_{lL}\,.
\label{001}
\end{equation}

We will assume that the flavor fields $\nu_{lL}$ (l=e, $\mu$, $\tau$)
are 
mixtures of the fields of neutrinos with definite masses 

\be
\nu_{lL} = \sum_{i=1}^{3}\,~ U_{li} \nu_{iL},
\label{002}
\ee

where $\nu_{i}$ is the field of neutrino (Dirac or Majorana) with mass $m_i$ and U is 3$\times$3 
PMNS  unitary matrix. 

If neutrino mass-squared differences  are so small that they can not be 
resolved in neutrino production and detection experiments, the state of the 
flavor neutrino $\nu_{l}$ with momentum $\vec p$ is given by the expression
\be
|\nu_{l}\rangle
= \sum_{i=1}^{3} U_{l i}^* \,~ |\nu_i\rangle \,.
\label{003}
\ee

Here $|\nu_i\rangle $ is the  state of neutrino with momentum $\vec p$,
and
energy 

\be
E_i = \sqrt{p^2 + m_i^2 } \simeq p + \frac{ m_i^2 }{ 2 p }; \,~
( p^2 \gg m_i^2 )
\label{004}
\ee

Let us assume that at t=0 in a weak decay flavor neutrino $\nu_{l}$ was produced.
At the moment t for the state vector we  have

\be
|\nu_{l}\rangle_{t}=\,~ e^{-iH_{0}\,t}\,~|\nu_{l}\rangle =
\sum_{i=1}^{3}\,U_{l i}^*\,e^{-iE_it}\,~|i\rangle,
\label{005}
\ee
where $H_{0}$ is the free Hamiltonian. 

Different mass components in (\ref{005}) have {\it different phases}. Thus,
the flavor content of the state $|\nu_{l}\rangle_{t}$ is different
from the flavor content of the initial state. For the probability of the transition $\nu_{l}\to \nu_{l'}$ from (\ref{005}) we have

\begin{equation}
{\mathrm P}(\nu_l\to \nu_{l'}) =
|\,~\delta_{l'l} +\sum_{i\geq 2} U_{l' i}  U_{l i}^*
\,~ (e^{- i \Delta m^2_{i 1} \frac {L} {2E}} -1)\,~|^2 \,,
\label{006}
\end{equation}
where $L\simeq t$ is the sorce-detector 
distance, 
$E$
is the neutrino energy and $\Delta m^2_{i 1} =  m^2_{i}- m^2_{1}$.
\footnote{ We label neutrino masses in such a way that $m_1 < m_2 < m_3.$}

Analogously, for the probability of the transition
$\bar\nu_{l} \to \bar \nu_{l'}$ we have
\begin{equation}
{\mathrm P}(\bar\nu_l \to \bar\nu_l') =
|\,~\delta_{l'l} +\sum_{i\geq2} U_{l' i}^*  U_{l i}
\,~ (e^{- i \Delta m^2_{i 1} \frac {L} {2E}} -1)\,~|^2 \,.
\label{007}
\end{equation}

We will make the following remarks. 

\begin{enumerate}

\item
In  the case of the  Dirac neutrinos the mixing matrix $U$ is characterized
by three mixing angles $\theta_{12}$, $\theta_{23}$ and $\theta_{13}$ and one  phase $\delta$.
We will be interested in the elements of the first row and the first
 column
of the matrix $U$. In the standard parameterization we have
\be
U_{e1} = \sqrt{ 1 -|U_{e 3}|^{2}}\,~ \cos\theta_{12};\,
U_{e2} = \sqrt{ 1 -|U_{e 3}|^{2}}\,~ \sin\theta_{12};\,
U_{e3} = \sin\theta_{13}
\,~e^{-i\,\delta}
\label{008}
\ee

For the $\mu,3$ and $\tau,3$ elements of the mixing matrix we have
\be
U_{\mu 3} = \sqrt{ 1 -|U_{e 3}|^{2}}\,~ \sin\theta_{23};\,~
U_{\tau 3} = \sqrt{ 1 -|U_{e 3}|^{2}}\,~ \cos\theta_{23}\,
\label{009}
\ee

In the case  of the Majorana neutrinos there are two additional phases 
in the matrix $U$. These  phases do not enter, however, into expressions 
for the transition probabilities (\ref{006}) and (\ref{007}).
Thus, in  the general case the transition probabilities of neutrinos and  antineutrinos 
depend on three angles, one  phase and two mass-squared differences
$\Delta m^{2}_{21}$ and $\Delta m^{2}_{32}$.

\item

Neutrino oscillations can be observed if for at  least  one 
$\Delta m^{2}$ the condition
$\Delta m^2 \frac {L} {E}\gtrsim 1 
$
is satisfied. In this inequality  $\Delta m^2 $
is the neutrino mass-squared difference in eV$^{2}$, $L$ is the 
 distance 
 in
m and $E$ is the neutrino energy in MeV.

\item

If CP invariance in the lepton sector holds, the matrix $U$ 
is real in the Dirac case:
\be
U=U^{*}
\label{011}
\ee

In the case of the Majorana neutrino $\nu_{i}$ the matrix $U$
satisfies the condition
\be
U_{l i}=U_{l i}^{*}\,~\eta_{i}\,,
\label{012}
\ee
where $\eta_{i}=\pm i$ is the CP parity of the Majorana neutrino $\nu_{i}$.

From (\ref{006}), (\ref{007}), (\ref{011}) and (\ref{012}) it follows that 
in the case of the CP invariance in the  lepton  sector
\be
{\mathrm P}(\nu_l \to \nu_{l'}) =
{\mathrm P}(\bar\nu_l \to \bar\nu_{l'}). 
\label{013}
\ee

\item

In the case of the mixing of two types of neutrinos the index $i$ in
(\ref{006}) and (\ref{007}) takes  only one  value $i=2$ and indexes $l$ and 
$l'$
take two values ($\mu, \tau$ or $\mu, e $ or $e, \tau$). For the  transition probability we have in this case

\be
{\mathrm P}(\nu_l \to \nu_{l'}) =
|\delta_{{l'}l} + U_{l' 2}  U_{l 2}^*
\,~ (e^{- i \Delta m^2 \frac {L} {2E}} -1)|^2 \,,
\label{014}
\ee
where $\Delta m^2= m^2_{2}-m^2_{1} $.
From this expression for the appearance probability
($l\not=l'$) we have
\begin{equation}
{\mathrm P}(\nu_l \to \nu_{l'}) =
{\mathrm P}(\nu_{l'} \to \nu_{l})=
\frac {1} {2} 
\,~ \sin^{2}2\,~\theta\,~
 (1 - \cos \Delta m^{2} \frac {L} {2E})\,,
\label{015}
\end{equation}
where 
$\theta$ is the mixing angle
($U_{l 2} = \sin\theta;\,~U_{l' 2} = \cos\theta$).

For the disappearance probability from the condition of the conservation of the probability we find
\be
{\mathrm P}(\nu_l \to \nu_l) 
=1-{\mathrm P}(\nu_l \to \nu_{l'})=
 1 - \frac {1}{2}\,~  \sin^{2}  2\theta \,~
(1 - \cos \Delta m^{2} \frac {L}{2E})\,.
\label{016}
\ee
Let us stress that in the case of the mixing of two types of  neutrinos the
following relations holds

\be
{\mathrm P}(\nu_l \to \nu_{l})=  
{\mathrm P}(\nu_{l'}  \to \nu_{l'})\,~~(l\not=l')
\label{017}
\ee
\item

All existing atmospheric neutrino oscillation data are perfectly  described 
if we  assume that $\nu_{\mu}\to \nu_{\tau}$ oscillations take place. 
From the analysis of the  Super-Kamiokande data  the following best-fit
values of the oscillations parameters were obtained \cite{S-K}
\be
\Delta m^{2}_{atm}=2.5\cdot 10^{-3}\rm{ eV}^{2};\,~\sin^{2}2 \theta_{atm}=1.0 
\,~(\chi^{2}_{\rm{min}}= 163.2/ 170\,\rm{d.o.f.}) 
\label{018}
\ee
The data of all solar neutrino experiments are well described
if we assume that 
 the probability of the solar
neutrinos to survive is given by the  two-neutrino MSW expression.
From the analysis of the data of all solar experiments for the best-fit 
 values of the oscillation parameters in the LMA region the following values were found
\cite{SNO}
\be
\Delta m^{2}_{\rm{sol}}=5\cdot 10^{-5}\rm{eV}^{2};
\,\tan^{2}\theta_{\rm{sol}}=0.34;\,~
\chi^{2}_{\rm{min}}= 57/72\, \rm{d.o.f.}.
\label{019}
\ee

Finaly the data of the KamLAND experiment \cite{Kamland}
are well described if we assume that the probability of the  reactor 
$\bar\nu_{e}$ is given by the two-neutrino
expression (\ref{016}). 
The best-fit values of the oscillation parameters 
\be
(\Delta m^{2})_{\rm{KamLAND}} = 6.9 \cdot 10^{-5}\rm{eV}^{2};\,~
(\sin^{2}2\,\theta)_{\rm{KamLAND}} =1\,
\label{020}
\ee
obtained from the analysis  of the  KamLAND data,
are compatible with the  solar LMA values (\ref{019}).
The  results of the  KamLAND experiment perfectly confirm effect of the oscillations of
electron neutrinos discovered in the solar neutrino experiments.  

\end{enumerate}
\section{Neutrino oscillations in the atmospheric and LBL experiments}

We will consider now vacuum neutrino oscillations  
in the atmospheric and long baseline (LBL) accelerator and reactor neutrino
experiments with $\frac{L}{E}\simeq 10^{-3}$ (see \cite{BGG}). 
 From analysis of  all existing neutrino oscillation data
it follows that  two-neutrino mass-squared differences $\Delta m^{2}_{sol}$ and $\Delta m^{2}_{atm}$  satisfy the inequality
$$\Delta m^{2}_{21}\simeq \Delta m^{2}_{sol}\ll \Delta m^{2}_{atm}
\simeq
\Delta m^{2}_{32}$$
Thus, for
$\frac{L}{E}\simeq 10^{-3}$ the phase  
$\Delta m^{2}_{21} \frac {L} {2E} $ in Eq.(\ref{006})       
and Eq.(\ref{007}) 
is small and the contribution of $\nu_{2}$
into the expressions  for the neutrino and antineutrino transition probabilities
can be neglected. We have

\be
{\mathrm P}(\nu_l \to \nu_{l'}) \simeq
|\,~\delta_{{l'}l} + U_{l' 3}  U_{l 3}^*
\,~ (e^{- i \Delta m^2_{32} \frac {L} {2E}} -1)\,~|^2 
\label{024}
\ee

From this expression for $l \not= l'$ we  have

\begin{equation}
{\mathrm P}(\nu_l \to \nu_{l'}) =
 \frac {1} {2} {\mathrm A}_{{l'};l}\,~
 (1 - \cos \Delta m^{2}_{32} \frac {L} {2E})\,
\label{025}
\end{equation}
where the oscillation amplitude is given by 
\be
{\mathrm A}_{{l'};l}= 4\,~|U_{l' 3}|^{2}\,~|U_{l 3}|^{2}={\mathrm A}_{{l};l'}
\label{026}
\ee

From (\ref{009}) and (\ref{026}) for the amplitudes of $\nu_{\mu}\to\nu_{\tau}$ and $\nu_{\mu}\to\nu_{e}$ transitions  we obtain
the following expressions
\be
 A_{\tau;\mu}= (1-|U_{e3}|^{2})^{2}\sin^{2}2\,\theta_{23};\,~
 A_{e;\mu}=4\,|U_{e3}|^{2}\,(1-|U_{e3}|^{2}\sin^{2}\,\theta_{23}
\label{027}
\ee

For the survival probability ${\mathrm P}(\nu_l \to \nu_l)$ 
from the condition of the conservation of the probability we have

\be
{\mathrm P}(\nu_l \to \nu_)= 1 -\sum_{l' \not= l}\,
 {\mathrm P}(\nu_l \to \nu_{l'}) =1 - \frac {1} {2}
{\mathrm B}_{l ; l}\,~ 
(1 - \cos \Delta m^{2}_{32} \frac {L} {2E})\,,
\label{028}
\ee
where

\be
{\mathrm B}_{l ; l}=\sum_{l' \not= l}\,
{\mathrm A}_{{l'};l}=
4\,~|U_{l 3}|^{2}
\,~(1 -|U_{l 3}|^{2}).
\label{029}
\ee

As it is seen from (\ref{025}) and (\ref{026}), the CP phase $\delta$ does
not enter into the expressions for the transition probabilities. 
This means that in the leading approximation 
 the relation
$$
{\mathrm P}(\nu_l \to \nu_{l'}) =
{\mathrm P}(\bar\nu_l \to \bar\nu_{l'}) $$ 
 is satisfied 
automatically.

Thus, the violation of the CP invariance in the lepton sector can be revealed
in neutrino oscillations only  if  the contribution of all three
massive neutrinos are involved.
Because the parameter
$\frac{\Delta m^{2}_{21}}{\Delta m^{2}_{32}}$ is small,  effects of the violation of the CP invariance in the lepton sector in the case  of the
 mixing of three massive neutrinos   are strongly suppressed.

The transition probabilities Eq.(\ref{025}) and Eq.(\ref{028})
in every channel have two-neutrino form. This is obvious consequence of the fact that we took into account only the major contribution of  the largest
neutrino mass-squared difference $\Delta m^{2}_{32}$.
Due to the unitarity  condition 
$\sum_{l}|U_{l 3}|^{2}=1 $ the transition probabilities 
Eq.(\ref{025}), Eq.(\ref{026}), Eq.(\ref{028}) and Eq.(\ref{029}) are
characterized by three parameters. We can choose 
$
\Delta m^{2}_{32},\,~\sin^{2}\theta_{23},\,~|U_{e3}|^{2}$.

The parameter $|U_{e3}|^{2} $ is  small.
This follows from the results of the  CHOOZ  and Palo Verde 
reactor neutrino experiments 
\cite{CHOOZ}.

No indications in favor of neutrino oscillations were found in these experiments.
For the probability of the reactor antineutrino to survive 
from (\ref{028})
we have
\be
{\mathrm P}(\bar \nu_e \to \bar \nu_e) =1 - \frac {1} {2}
{\mathrm B}_{e; e}\,~ 
(1 - \cos \Delta m^{2}_{32} \frac {L} {2E})\,,
\label{031}
\ee
where 
\be
{\mathrm B}_{e ; e} =4\, |U_{e 3}|^{2}\,(1-|U_{e 3}|^{2})
\label{032}
\ee

From exclusion plots obtained from the analysis of the CHOOZ and Palo Verde data we have
\be
{\mathrm B}_{e ; e} \leq {\mathrm B}_{e ; e}^{0},
\label{033}
\ee
where the upper bound ${\mathrm B}_{e ; e}^{0}$
depends on 
$\Delta m^{2}_{32}$.
From  Eq.(\ref{032}) and Eq.(\ref{033}) we obtain the bound \footnote {The large values
of  $|U_{e 3}|^{2}$ ($
|U_{e 3}|^{2} \geq
1-\frac{1}{4}\, {\mathrm B}_{e ; e}^{0}$)
are excluded by the solar neutrino  data.}

\be
|U_{e 3}|^{2} \leq
\frac{1}{4}\,  {\mathrm B}_{e ; e}^{0}
\label{034}
\ee
For the S-K best-fit value 
$\Delta m^{2}_{32}=2.5\cdot 10^{-3}\rm{ eV}^{2}$
from the CHOOZ exclusion plot we have 

\be
|U_{e 3}|^{2}\leq 4\cdot 10^{-2}\,~ (95 \%\,~\rm{CL}).
\label{034}
\ee

Taking into account accuracies of the present day experiments 
we can neglect 
$|U_{e 3}|^{2}$ in the expressions for the transition probabilities. In this approximation $\Delta m^{2}_{32}\simeq \Delta m^{2}_{\rm{atm}}$ and neutrino oscillations in the atmospheric range of $ \Delta m^{2}$ are  pure
vacuum two-neutrino $\nu_{\mu}\to\nu_{\tau}$ oscillations. The S-K and other atmospheric neutrino data are in agreement with such  picture.

\section{Neutrino oscillations in the solar and KamLAND experiments}

In the framework of the three-neutrino mixing we will consider now
 neutrino oscillations in the solar range of $\Delta m^{2}$ 
(see \cite{BGG}).
The $\nu_{e}$ ($\bar\nu_{e}$) vacuum survival probability can be written in the  form

\be
{\mathrm P}(\nu_e\to\nu_e)={\mathrm P}(\bar\nu_e\to\bar\nu_e)
=
\left|
\sum_{i=1, 2}| U_{e i}|^2 \, 
 e^{ - i
 \, 
\Delta{m}^2_{i1} \frac {L}{2 E} }
 + | U_{e 3}|^2  \, 
 e^{ - i
 \, 
\Delta{m}^2_{31} \frac {L}{2 E} }\,\right|^2
\label{035}
\ee
We are interested in  the survival probability averaged over
the region where neutrinos are produced, over neutrino spectrum,  energy resolution etc. 
Because of the inequality $\Delta m^{2}_{31}\gg \Delta m^{2}_{21}$, 
in the expression for the averaged  probability
the interference between the first and the second 
terms of (\ref{035})
disappears. The  averaged survival probability can be presented in the form

\be 
{\mathrm P}(\nu_{e}\to\nu_{e})=
=|\,U_{e 3}|^{4}+ (1-|U_{e 3}|^{2})^{2}\,~
P^{(1,2)}(\nu_{e}\to\nu_{e})\,.
\label{036}
\ee

Here 

\begin{equation}
{\mathrm P}^{(1,2)}(\nu_e \to \nu_e) =
 1 - \frac {1} {2}\,~A^{(1,2)}\,~ 
(1 - \cos \Delta m^{2}_{21} \frac {L} {2E})\,,
\label{037}
\end{equation}

where
\be
A^{(1,2)}= 4\,\frac{ |U_{e 1}|^{2}\,|U_{e 2}|^{2}} 
{( 1-   |U_{e 3}|^{2})^{2} }=\sin^{2}2\,\theta_{12}
\label{038}
\ee

The expression (\ref{036}) is also valid in the case
of matter. In this case
$P^{(1,2)}(\nu_{e}\to\nu_{e})$ is the two-neutrino $\nu_{e}$ survival
probability in matter, calculated under the condition that the
density of electrons $\rho_{e}(x)$ in the effective
Hamiltonian of the interaction of neutrino with matter is
changed by $(1-|U_{e 3}|^{2})\,\rho_{e}(x)$.

Thus, $\nu_{e}$ ($\bar\nu_{e}$) survival probability is characterized in the solar range of
$\Delta m^{2}$ by three parameters
$
\Delta m^{2}_{21},\,~\tan^{2}\theta_{12},\,~|U_{e3}|^{2}$
. 
The only common parameter for the atmospheric and solar ranges of 
$\Delta m^{2}$ is $|U_{e3}|^{2}$. In the approximation 
$|U_{e3}|^{2} \to 0$ oscillations in the solar range of $\Delta m^{2}$
are described by the two-neutrino expression

\be 
{\mathrm P}(\nu_{e}\to\nu_{e})=
=
P^{(1,2)}(\nu_{e}\to\nu_{e})
\label{040}
\ee

and $\Delta m^{2}_{21}\simeq \Delta m^{2}_{\rm{sol}}\,~
\tan^{2}2\,\theta_{12}\simeq  \tan^{2}2\,\theta_{\rm{sol}}$.

Thus, in the  leading approximation 
($|U_{e3}|^{2} \to 0$, $\Delta m^{2}_{21}\,~\frac{L}{2E}\to 0$
in the atmospheric and LBL experiments) neutrino oscillations in the 
atmospheric and solar ranges of $\Delta m^{2}$ are decoupled (see \cite{BGG}).

\section{Neutrinoless double $\beta$-decay}

The search for neutrinoless double $\beta$-decay 
($(\beta\beta)_{0\nu}$ -decay)
\be
(A,Z) \to (A,Z+2)+ e^{-}+ e^{-}
\label{041}
\ee
of some even-even nuclei is the most sensitive and direct way of the investigation of the nature of neutrinos with definite masses $\nu_{i}$:
 the process (\ref{041})
is allowed only if  
massive neutrinos  are Majorana particles.

The  matrix element of the $(\beta\beta)_{0\nu}$ -decay is the product of the  effective Majorana mass

\be
m_{\beta\beta} = \sum_{i}U^{2}_{ei}\,m_{i}.
\label{042}
\ee

and nuclear matrix element. 
Taking into account different calculations of the nuclear matrix elements,
for the effective Majorana mass from the data of the most precise 
$^{76} \rm{Ge}$ experiments  \cite{HM}
the following upper bound was obtained
\be
|m_{\beta\beta}| \leq (0.3-1.2)\,~\rm{eV}\,. 
\label{043}
\ee
Many new experiments on the search for the neutrinoless 
double $\beta$-decay are in preparation at present \cite{ndm03}.
In these experiments the sensitivities
\be
|m_{\beta\beta}| \simeq (1\cdot10^{-1}-1.5\cdot 10^{-2})\,~\rm{eV}
\label{044}
\ee
are expected to be achieved.

The evidence for neutrinoless double $\beta$-decay would be a proof
that neutrinos $\nu_{i}$
are Majorana particles.
We would like to stress here  that the value of the effective Majorana mass $|m_{\beta\beta}|$,  
combined with the values of the neutrino oscillation parameters,
would enable us to obtain an information about the character of
the neutrino mass spectrum, minimal neutrino mass $m_1$ and 
possibly Majorana CP phase (see\cite{SP}).
Let us  consider three typical
neutrino mass spectra.

\begin{enumerate}

\item
The hierarchy of neutrino masses $m_1 \ll m_2\ll m_3$.

We have in this case the following upper bound 

\be
|m_{\beta\beta}| \leq \sin^{2} \theta_{\rm{sol}}\,\sqrt{\Delta m^{2}_{\rm{sol}}}+ 
|U_{e3}|^{2}\,\sqrt{\Delta m^{2}_{\rm{atm}}}.
\label{045}
\ee

Using the best-fit values of the oscillation parameters 
and the CHOOZ bound on $|U_{e3}|^{2}$ (see (\ref{018}), (\ref{019}) and 
(\ref{034})),
we obtain the bound
\be
|<m>| \leq 3.8 \cdot 10^{-3}\rm{eV}\,,
\label{046}
\ee

which is significantly smaller than the expected sensitivities 
of the future $(\beta \beta)_{0\nu}$- experiments.

\item
Inverted hierarchy of neutrino masses: $m_{1}\ll m_{2}<m_{3}$.

The effective Majorana mass is given in this case by the expression

\be
|m_{\beta\beta}|\simeq  \left( 1 - \sin^{2}2\,\theta_{\rm{sol}}\,\sin^{2}\alpha \right)
^{\frac{1}{2}}\,\sqrt{\Delta m^{2}_{\rm{atm}}},
\label{047}
\ee
where $\alpha=\alpha_{3}-\alpha_{2}$ is the the difference of the Majorana 
CP phases 
of the elements $U_{e3}$ and $U_{e2}$.

From this expression it follows that

\be
\sqrt{\Delta m^{2}_{\rm{atm}}}
\,~ |\cos2\,\theta_{sol}|\lesssim |m_{\beta\beta}|
\lesssim \sqrt{\Delta m^{2}_{\rm{atm}}}\,,
\label{048}
\ee
where the upper and lower bounds correspond to the 
case of the CP conservation with the equal and opposite CP parities 
of $\nu_{3}$ and $\nu_{2}$.

Using the best-fit value of the parameter $\tan^{2}\theta_{sol}$
(see (\ref{019})), we have

\be
\frac{1}{2}\,\sqrt{\Delta m^{2}_{\rm{atm}}}
\lesssim |m_{\beta\beta}|
\lesssim\sqrt{\Delta m^{2}_{\rm{atm}}}\,,
\label{049}
\ee

Thus, in the case of the inverted mass hierarchy
the scale of $|m_{\beta\beta}|$ is determined
by $\sqrt{\Delta m^{2}_{\rm{atm}}}$.
If the value of $|m_{\beta\beta}|$ is in the range (\ref{049}),
which can be reached in the future experiments 
it would be an argument in favour of inverted neutrino mass hierarchy.

\item

Practically degenerate neutrino mass spectrum
$m_2 \simeq m_3\simeq m_1\gg \sqrt{\Delta m^{2}_{\rm{atm}}}$.

For the effective Majorana mass
we have in this case
\be
|m_{\beta\beta}| \simeq m_{1}\,~|\sum_{i=1}^{3}U_{ei}^{2}|.
\label{050}
\ee
Neglecting small contribution of
$|U_{e3}|^{2}$ ),
for $|m_{\beta\beta}||$ we obtain the relations (\ref{047})-(\ref{049}), 
in which 
$\sqrt{\Delta m^{2}_{\rm{atm}}}$ must be changed by $ m_1$. 
For neutrino mass $m_{1}$ we have in this case the range

\be
|m_{\beta\beta}| \leq m_{1}\leq\frac{|m_{\beta\beta}|}
{|\cos 2\,\theta_{\rm{sol}}|}\simeq 2\,|m_{\beta\beta}|
\label{051}
\ee
The parameter
$\sin^{2}\alpha$,
which characterize the violation of the CP invariance in the lepton sector,
is given by the following relation

\be
\sin^{2}\alpha \simeq \left( 1 -\frac{|m_{\beta\beta}|^{2}}{m_{1}^{2}}
\right)\,~\frac{1}{\sin^{2}2\,~\theta_{\rm{sol}}}\,.
\label{052}
\ee

If the mass $m_1$ is measured in the future $\beta$-decay
experiments \cite{Katrin}
and the value of the parameter $\sin^{2}2\,~\theta_{\rm{sol}}$ 
is determined  in the solar and
KamLAND  experiments,
from the results of the future $(\beta \beta)_{0\nu}$- experiments  
an information on the Majorana
CP phase can be inferred.

\end{enumerate}

The determination of the
Majorana mass $|m_{\beta\beta}|$ from the measurement of the life-time
of the $(\beta \beta)_{0\nu}$-decay
 requires the knowledge of the nuclear matrix elements.
At present there are large uncertainties in the calculation of these
quantities :
different calculations of the
lifetime of the $ (\beta \beta)_{0\,\nu}$-decay differ
by about one order of magnitude.

If $(\beta \beta)_{0\nu}$-decay of
{\em different nuclei} will be observed, from the ratios  of the life-times 
(which depend on the ratios  of the nuclear matrix elements and known space factors) model independent conclusions
on different calculations of the nuclear matrix elements can be inferred 
\cite{BGr}.

\section{Conclusion}

There are many unsolved problems in the physics of massive and mixed neutrinos.
From our point of view the most urgent ones are the following: 
\begin{itemize}

\item

What is the value of the parameter $|U_{e3}|^{2}= \sin^{2}\,\theta_{13}$?

The answer to this question possibly will be obtained in LBL accelerator
experiments searching for $\nu_{\mu} \to \nu_{e}$ oscillations and LBL reactor
experiments searching for disappearance of reactor $\bar\nu_{e}$ 
\cite{ndm03}.
\item

What is the nature of neutrinos with definite masses? Are they Dirac or Majorana particles?

The answer to this question possibly will be obtained in future experiments on
the search for neutrinoless double $\beta$-decay\cite{ndm03}.

\item

What is the value of the minimal neutrino mass $m_{1}$?

 The answer to this question possibly will be obtained
in future tritium experiment KATRIN \cite{Katrin}  and/or from cosmology 
\cite{ndm03}.

\end{itemize}

\end{document}